\documentclass{article}

\usepackage[final,nonatbib]{tackling_climate_workshop_style}

\usepackage{wrapfig}
\usepackage[utf8]{inputenc} 
\usepackage[T1]{fontenc}    
\usepackage{hyperref}       
\usepackage{url}            
\usepackage{booktabs}       
\usepackage{amsfonts}       
\usepackage{nicefrac}       
\usepackage{microtype}      
\usepackage{graphicx}
\usepackage{biblatex} 
\usepackage{subcaption}
\title{Unlocking the Potential of Renewable Energy Through Curtailment Prediction}

\author{Bilge Acun$^{1}$, Brent Morgan$^{1}$, Carole-Jean Wu$^{1}$, Henry Richardson$^{2}$, Nat Steinsultz$^{2}$\\ \\
$^{1}$Meta \hskip 1em $^{2}$WattTime}

\begin{document}

\maketitle

\vspace{-0.5cm}
\begin{abstract}
A significant fraction (5-15\%) of renewable energy generated goes into waste in the grids around the world today due to oversupply issues and transmission constraints. Being able to predict \textit{when} and \textit{where} renewable curtailment occurs would improve renewable utilization. The core of this work is to enable the machine learning community to 
help decarbonize electricity grids
by unlocking the potential of renewable energy through curtailment prediction.      
\end{abstract}

\vspace{-0.5cm}
\section{Introduction}
\vspace{-0.2cm}
Decarbonizing electricity grids reliably and cost-effectively is critical towards a clean energy future. Significant investment into renewable energy infrastructure from the past decade has dramatically increased the amount of wind and solar deployed on grids around the world from 630 TWh in 2012 to over 3,400 TWh in 2022~\cite{energy_institute}.
However, as renewable generation increases, the amount of energy curtailed due to oversupply during certain hours also increases.

\begin{wrapfigure}{r}{0.5\textwidth}
  \begin{center}
  \includegraphics[width=0.50\textwidth]{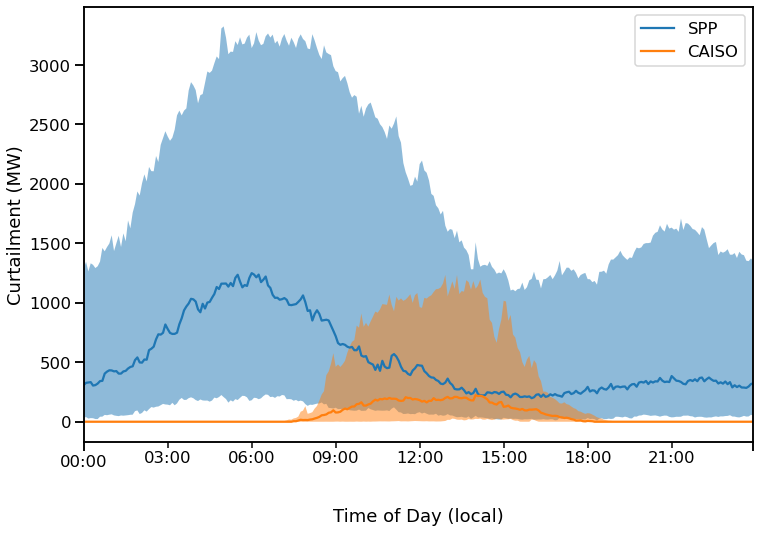}
  \end{center}
  \vspace{-0.4cm}
  \caption{Median (solid line) and 25\%-75\% range (shaded) of curtailment values, aggregated by time of day, for SPP (primarily wind curtailment) and CAISO (primarily solar curtailment).}
  \label{fig:curtailment_distribution}
\end{wrapfigure}

As intermittent renewable energy generation becomes increasingly prevalent, the risk of curtailment grows significantly~\cite{prol2023no}. Curtailment is a key challenge facing the world where power grids are decarbonized. For example, 5-15\% of the renewable energy generated is curtailed in Europe and China respectively~\cite{bunodiere2020renewable}. In the US, in 2022, SPP curtailed over 11,000 GWh of wind energy, which was more than 10\% of all wind energy produced in the region~\cite{spp_state_of_market}. CAISO curtailed almost 2,450 GWh of renewable energy which was almost 7\% of all solar energy produced~\cite{caiso_curtailment}, as illustrated in Figure~\ref{fig:curtailment_distribution}. The curtailment is happening frequently as well, occurring at some location in CAISO over 30\% of all time steps in 2021, and over 60\% of all times in ERCOT and SPP. The frequent occurrence of curtailment means that there is opportunity for greater emissions reductions through careful load management. 

Historically, dispatchable fossil resources were matched to electricity demand from consumers. We now have the opportunity to reverse that paradigm and match demand to variable renewable energy supply to sop up the excess clean energy. Flexible loads, like electric vehicle charging, hot water heating with heat pumps, and delay-tolerant computation in datacenters~\cite{ acun2023carbon, lin2023adapting, radovanovic2022carbon} can be scheduled to take advantage of this low emissions electricity. For example, managed EV charging can reduce associated emissions by up to 65\%~\cite{rmi}. But to take advantage of this curtailed renewable energy, we need to \textit{accurately predict when and where curtailment occurs} — a challenging problem due to the intermittent nature of renewable energy production.   

\vspace{-0.2cm}
\section{ML for Curtailment Prediction}
\vspace{-0.2cm}

Understanding when and where renewable energy curtailment occurs provides a massive opportunity for emissions reductions.
When renewable energy is oversupplied, or is constrained by transmission capacities, it is curtailed. If users could increase load at the times and near places where curtailment occurs, electricity usage is increased without the associated emission overhead, having near-zero marginal emissions during that time. For example, managed EV charging can reduce associated emissions by up to 65\%~\cite{rmi}.
In addition to lowering carbon emissions, reducing curtailment increases revenues for renewable energy owners and reduces the overall system costs for grid consumers. As renewable development continues, the opportunity to reduce emissions through load flexibility will increase in magnitude, frequency and location. Temporal and spatial granularity of the models are crucial to \textit{detect} and \textit{forecast} the curtailments accurately.

\textbf{Temporal Granularity:} Curtailment can vary significantly from one 5-minute settlement period of the Independent System Operators (ISO)\footnote{ISO is the entity that manages dispatch of generators in the grid in most regions of the US~\cite{rff} to the next.
}. See Appendix Figure~\ref{fig:curtailment_minutes} for an example five minute-level time series and Figure~\ref{fig:curtailment_time_series} for an example of seasonal trends. This means that a successful curtailment prediction model needs to operate at a granularity finer than hourly.

\textbf{Spatial Granularity:} While curtailment can be a widespread phenomenon, it is most frequently localized to a subsection of the ISO grid due to transmission constraints.
This means that the geographic granularity needs to be smaller than at the ISO level. 
Fortunately, there is a growing resource of ground-truth data about curtailment available now. Many ISOs publish ground truth curtailment data with varying time delays, typically 30 to 90 days, although some may be as soon as a day or up to a year delay. The data can have different levels of spatial granularity. Table~\ref{tab:iso_table} summarizes the characteristics of the ISO-specific datasets. Unfortunately not all ISOs provide nodal level granularity. However we propose using locational marginal pricing (LMP) data to serve as a replacement for granular ground truth data.


\subsection{Detecting Curtailment Using Nodal Locational Marginal Pricing (LMP) Data}
\begin{figure}
\centering
\begin{subfigure}{.5\textwidth}
  \centering
  \includegraphics[width=1\textwidth]{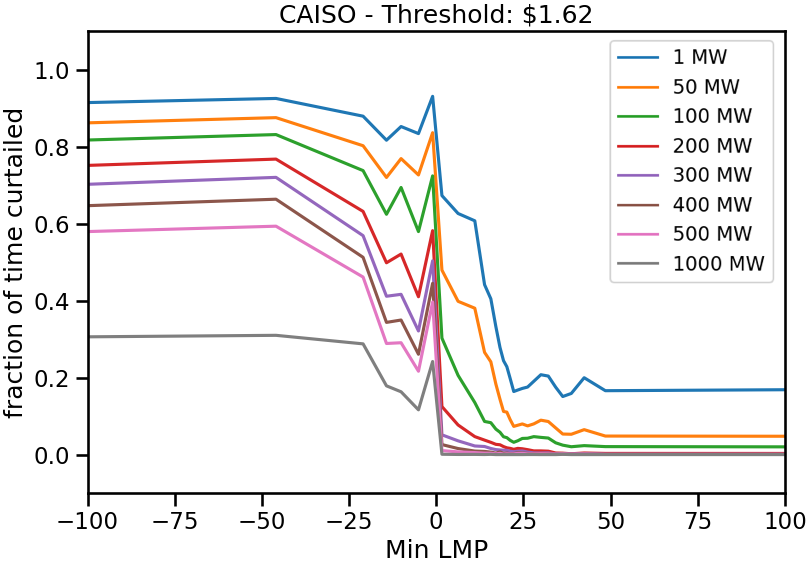}
  \label{fig:sub1}
\end{subfigure}%
\begin{subfigure}{.5\textwidth}
  \centering
  \includegraphics[width=.7\textwidth]{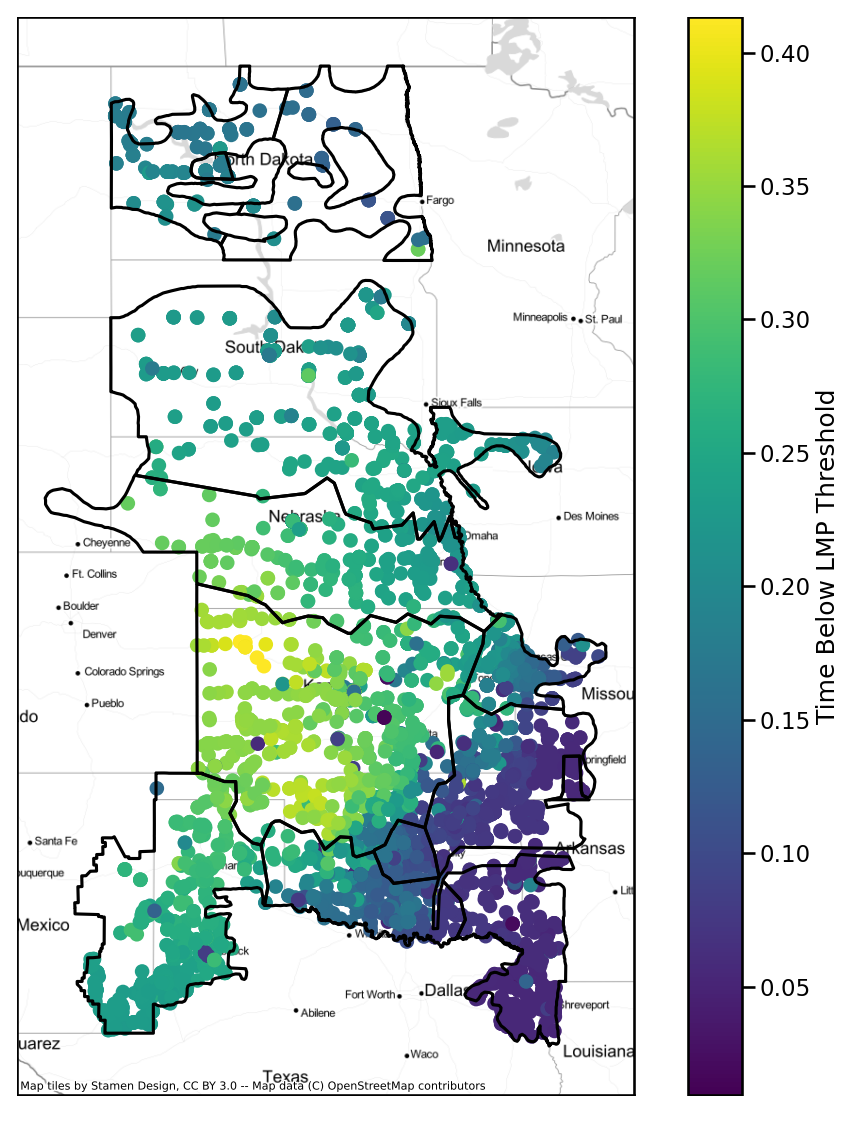}
  \label{fig:sub2}
\end{subfigure}
\caption{Left: Curtailment frequency as a function of the minimum LMP for different curtailment amounts (MW). Right: Heatmap of the times below LMP threshold at nodal level in SPP.}
\label{fig:lmp}
\end{figure}

While detailed information of where curtailed energy is not always available, many ISOs provide detailed nodal locational marginal pricing (LMP) data realtime. Not all markets globally use nodal LMP for dispatch, however all major ISOs in the US have moved to a nodal LMP structure. Since the majority of curtailment in current markets is driven by economic dispatch, LMP provides useful information about where curtailment is addressable. Renewable generators typically bid at zero or negative prices, so it is reasonable to assume that when nodal LMP crosses below a certain threshold, renewable energy is being curtailed or dispatched to meet marginal changes in demand. Figure~\ref{fig:lmp}-left depicts curtailment frequency as a function of the minimum LMP for CAISO, showing a 50\% curtailment likelihood at an LMP of \$1.62. The curtailment threshold may be ISO and condition dependent. Figure~\ref{fig:lmp}-right illustrates the significantly varying curtailment that occurs in various sub-regions of SPP based on the LMP threshold. This underscores the importance of curtailment prediction at the node granularity. In some ISOs, such as MISO, negative LMP values can be driven by imports or other types of transmission constraints, which are not related to curtailment.

Since curtailment is a localized phenomenon, it is important that forecasts of curtailment also be localized. For some use cases, such as grid battery storage, the ability to reduce emissions depends on accurately forecasting the behavior of individual pricing nodes. Each ISO can have thousands of unique pricing nodes and because curtailment is driven by extreme shifts in local dispatch, using averaged LMP values can miss many instances of localized curtailment. Existing work has been done on forecasting curtailment and marginal emissions, but not at a nodal level~\cite{hadiandeep, wangforecasting}.

\subsection{Forecasting Curtailment}
Once we have a reasonable model of when and where curtailment is occurring, accurate forecasting of those values is critical to driving real-world impact through load-shifting technologies.

\textbf{Forecasting window:} Software-controlled devices could schedule when they use electricity based on a curtailment forecast. If the model were used to control real-time load, it has to produce predictions in real-time, using real-time data, as well as short term forecasts to inform load management decisions. Many load shifting technologies, like grid-scale batteries, have to make decisions some time ahead due to market structures or technological constraints. In order to determine the best time to consume electricity, current conditions need to be compared to what is expected over the entire window of use. This window can be different for different technologies — low-priority tasks may be able to delay execution for up to a week, a grid-connected battery may have a 24-hour window in which it can charge and discharge, a residential EV may have an 8-hour overnight window for charging, and a smart thermostat may only have a 60 minute window in which it can shift usage before it impacts home temperatures. 
For devices that wish to participate in electricity markets, 95\% of the energy transactions are done in a day ahead markets and is based on the load forecasts of the next day~\cite{rff}. Therefore, 24-hour prediction of the forecasts of curtailment would be the most useful window for majority of the applications.

\textbf{Forecasting signal type:} Curtailment forecasting can be handled as either a regression or classification problem, depending on the data available and the application of the forecasts. Some ISOs only report a boolean value as to whether or not curtailment may be occurring, instead of the total amount of curtailment in MWh. For many load-shifting technologies, it may not matter the total amount of curtailed energy, but simply whether it is above or below some threshold relevant to the application. For example, it may not matter for the operation of a 50MW battery whether curtailment is 1,000MW or 10,000MW at a given time, simply that there is sufficient curtailed energy at that time. Thus in some cases a binary or binned signal may be sufficient for supporting load-shifting applications.

\vspace{-0.2cm}
\section{Call to Action}
\vspace{-0.2cm}
Curtailment and emissions-informed load-shifting technology has the potential to reduce real-world operating emissions today in many locations, which will rapidly increase in the future.
To make this possible,
we will need accurate data and forecasts about curtailment. Forecasting \textit{granularity, window} and \textit{the signal type} are important properties to consider to make a machine learning model high quality and useful. Existing models today lack these properties.
Until more ISOs regularly publish locationally granular real-time curtailment data, nodal LMP based models can be used to detect where curtailed energy is likely available.
To support this, we have collected, cleaned, and standardized historical curtailment data and LMP for the ISOs in North America. As part of this call to action, we have made some curtailment and nodal level LMP data available along with resources on where more data can be accessed, depending on ISO specific data licenses~\cite{nat_steinsultz_2023_10235806}.

Accelerating power grid decarbonization is a key step towards a clean energy economy, where renewable is the fastest growing energy source. A significant amount of renewable energy is stranded in the grids today because of oversupply and transmission congestion. To realize the full potential of renewable energy, we need a coordinated approach:

\begin{itemize}
    \item This paper takes the first step by making compiled datasets of nodal LMP values available for several ISOs. This enables curtailment prediction using machine learning models.
    \item We need grid operators to provide more detailed information publicly to give consumers signals about where curtailed energy is available in order for power consumers to improve curtailment prediction models. These signals then will enable load shifting decisions by power consumers.
    \item We need even more technologies to support carbon-aware load shifting to reduce emissions in increasingly variable grids. To accomplish this, we will need carbon-aware software standards, such as the carbon-aware scheduler demonstrated by Low-Carbon Kubernetes Scheduler~\cite{james}, the development of a carbon aware SDK by the Green Software Foundation~\cite{GSF} or carbon-aware demand response for large-scale datacenters by Carbon Responder~\cite{xing2023carbon}.
\end{itemize}

\begin{table}[ht]
\fontsize{9pt}{9pt}\selectfont
\begin{tabular}{lccc}
\hline
\\[-0.5em]
\textbf{ISO} & \textbf{Granularity} & \textbf{Reported Data} &  \textbf{\% Time with Curtailment} \\ \hline
\\[-0.5em]
SPP & 5 min & System wide curtailed power & 47.4 \\ \hline
\\[-0.5em]
CAISO & 5 min & System wide curtailed power & 23.1 \\ \hline
\\[-0.5em]
NYISO & Hourly & System wide curtailed power & 7.3 \\ \hline
\\[-0.5em]
PJM & Hourly & Percent of nodes with marginal fuel & 31.4 \\ \hline
\\[-0.5em]
MISO & Hourly & Regonal marginal fuel flag & 19.3 \\ \hline
\\[-0.5em]
ISONE & Hourly & System wide marginal fuel flag & 23.7 \\ \hline
\\[-0.5em]
ERCOT & 5 min & Plant output capability and actual output & 42.3 \\ \hline
\\[-0.5em]
IESO & Hourly & Plant output capability and actual output & 35.2 \\ \hline
\\
\end{tabular}
\caption{Historical curtailment information provided by the ISOs.}
\label{tab:iso_table}
\end{table}

\printbibliography
\appendix

\section{Supplementary Figures}

\begin{figure}[thb!]
  \centering
  \includegraphics[width=0.6\textwidth]{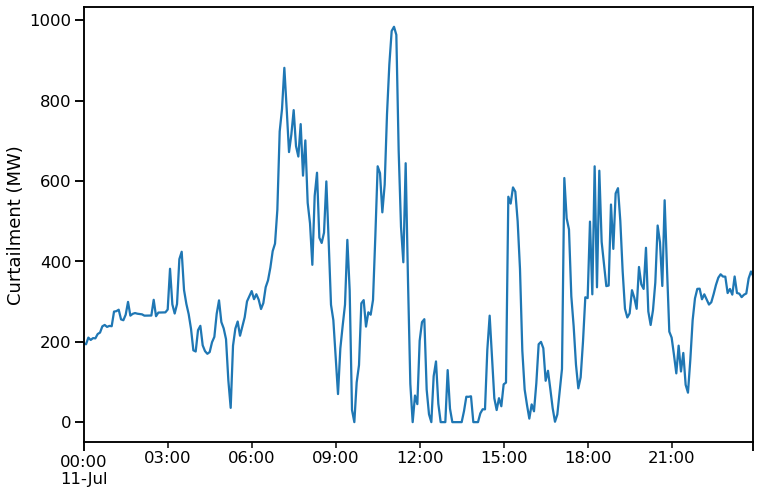}
  \caption{SPP wind curtailment in 5-minute steps on a random day.}
  \label{fig:curtailment_minutes}
\end{figure}

\begin{figure}[thb!]
  \centering
  \includegraphics[width=0.6\textwidth]{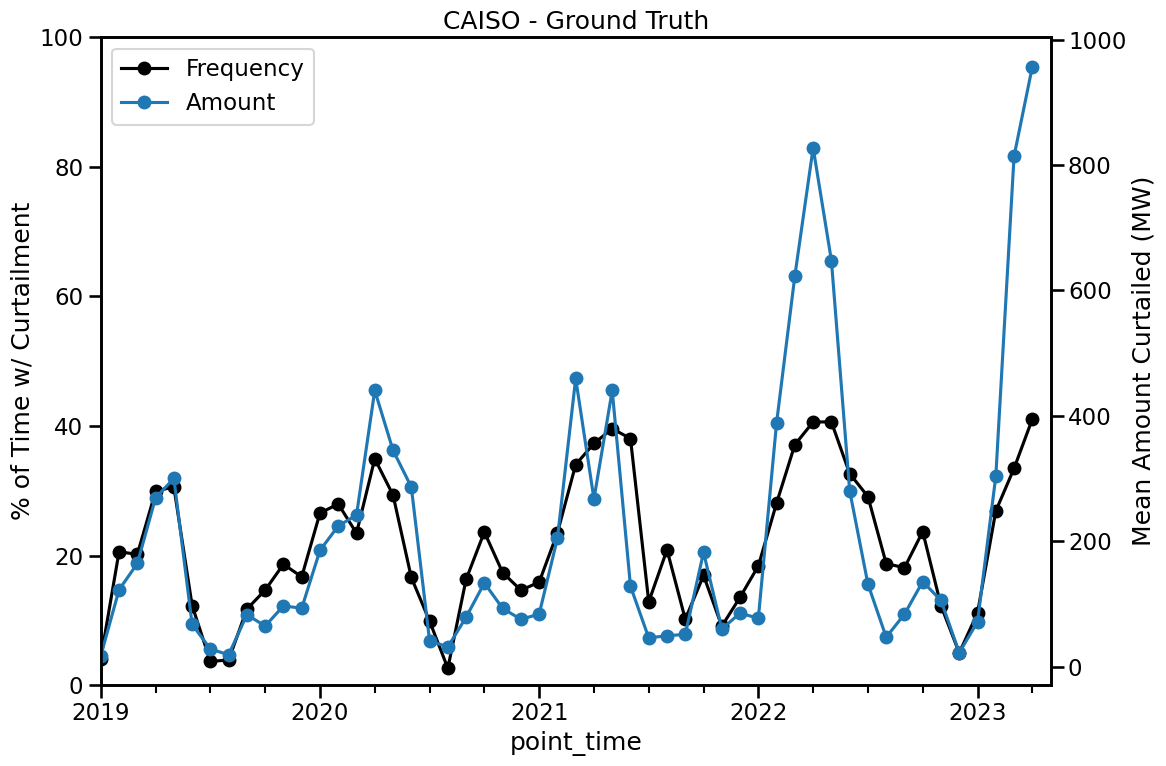}
  \caption{CAISO curtailment frequency and amount shown as time series from 2019-2023.}
  \label{fig:curtailment_time_series}
\end{figure}

\section{Forecasting Metrics}
\begin{figure}[t]
  \centering
  \includegraphics[width=0.6\textwidth]{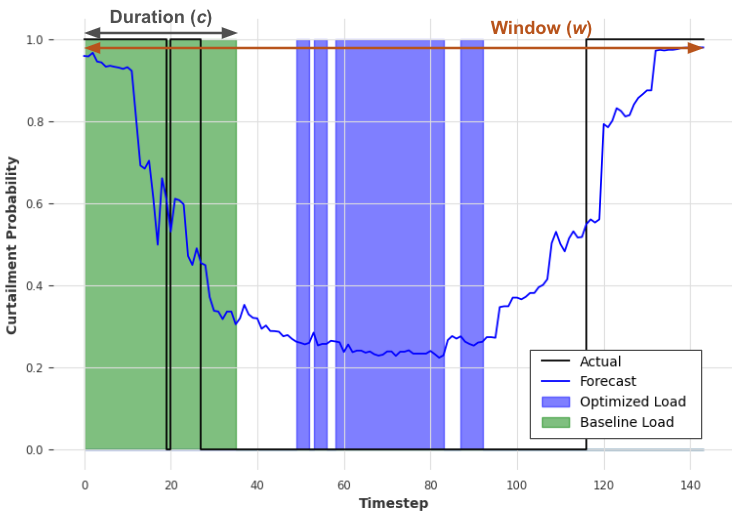}
  \caption{Illustration of load shifting forecast metrics.}
  \label{load_shifting}
\end{figure}

While standard error metrics are useful in evaluating curtailment forecasting models, the most important measure to drive real-world impact is correctly selecting times of curtailment during a fixed window when load is available. For a load-shifting technology which needs to use c minutes of energy during a window of length w starting at time $t$, the impact of using a forecast can be evaluated as the mean curtailment value during the c lowest minutes of the forecast during $t$ to $t+w$. This impact can be compared to the baseline of either immediate usage during the first $c$ minutes of the window w, or random-time usage, which is equivalent to the mean curtailment value during $t$ to $t+w$. Figure~\ref{load_shifting} provides an illustration of load shifting highlighting the importance these metrics.

\section{Additional Datasets}
There are several additional external data sets that may be useful in detecting and forecasting curtailment. The EIA makes hourly demand, generation and interchange data available at the ISO region/sub-region level historically real-time through their API~\cite{eia_api}. Many ISOs also independently make data directly available through an API, including 5-minute demand and generation data, along with additional datasets such as transmission binding constraints, that may be valuable in addressing both the temporal and spatial granularity needed for this problem. Given that renewable generation is driven by meteorological conditions, historical weather forecasts, such as those produced by NOAA's GFS model~\cite{noaa_gfs} are also important.

\end{document}